\documentclass[pre,aps,twocolumn,showpacs]{revtex4}
\usepackage{graphicx,amsmath,amsfonts,slashbox}
\usepackage[usenames]{color}
\usepackage{amssymb}
\usepackage{amsthm}
\input{epsf}
\begin{document}
\title{Extinction risk and structure of a food web model}

\author{Andrzej P\c{e}kalski}
\affiliation {Institute of Theoretical Physics, University of Wroc{\l}aw,\\
pl. M. Borna 9, 50-204 Wroc{\l}aw, Poland}
\email{apekal@ift.uni.wroc.pl}
\author{Janusz Szwabi\'nski}
\affiliation {Institute of Theoretical Physics, University of Wroc{\l}aw,\\
pl. M. Borna 9, 50-204 Wroc{\l}aw, Poland and D\'epartement de Physique
Th\'{e}orique, Universit\'{e} de Gen\`{e}ve, quai E. Ansermet 24,
1211 Gen\`{e}ve 4, Switzerland}
\email{Janusz.Szwabinski@physics.unige.ch}
\author{Ioana Bena}
\email{Ioana.Bena@physics.unige.ch}
\author{Michel Droz}
\email{Michel.Droz@physics.unige.ch}
\affiliation {D\'epartement de Physique Th\'{e}orique, Universit\'{e} de
Gen\`{e}ve, quai E. Ansermet 24, 1211 Gen\`{e}ve 4, Switzerland}

\begin{abstract}
We investigate in detail the model of a trophic  web proposed by
Amaral and Meyer [Phys. Rev. Lett. {\bf 82}, 652 (1999)]. We focused
on small-size systems that are relevant for real biological food webs and
for which the fluctuations are playing an important role.
We show, using Monte Carlo simulations, that such
webs can be non-viable, leading to extinction of all species in
small and/or weakly coupled systems. Estimations of the extinction times
and survival chances are also given. We show that before the extinction
the fraction of highly-connected species (``omnivores") is increasing.
Viable food webs exhibit a pyramidal structure, where the density of occupied
niches is higher at lower trophic levels, and moreover the occupations of
adjacent levels are closely correlated. We also demonstrate  that the
distribution of the lengths of food chains has an exponential
character and changes  weakly with the  parameters of the model. On the
contrary, the distribution of avalanche sizes of the extinct species
depends strongly on the connectedness of the web. For rather loosely
connected systems we recover the power-law type of behavior with
the same exponent as found in earlier studies, while for
densely-connected webs the distribution is not of a power-law type.
\end{abstract}


\maketitle

\section{Introduction}

Food webs describe the resources and trophic relationships among species within an ecosystem.
The first semi-quantitative descriptions of food webs were given by biologists
at the end of the nineteen century~\cite{paine,cohen}. Later on prey-predator relationship between
species were defined in terms of oriented graphs with hierarchical or layered
structures~\cite{pimm}.
The problem of describing such food webs was then taken over by mathematicians
and physicists, and different modeling levels and types of models
have been proposed.

A first group of models is constituted by the so-called
{\em static models} in which the links between different
species are assigned once and for all, according to different scenarios
(random, scale-free or small-world graphs~\cite{watts,barabasi}, for example).
Some properties
of these food webs were analyzed and compared
with available biological data,
and the comparison usually turned out to be quite poor.

The second group of models contains the so-called
{\em dynamic food web models}. The
novelty consists in recognizing that the links between
the species are generally not
arbitrary and quenched, but emerge as the result of
some intrinsic biological dynamics.
There are then many possibilities to model the
evolutionary dynamics~\cite{DmcK}. The simplest
one concerns two-layered systems with prey-predator
Lotka-Volterra type of dynamics
(for a short review, see~\cite{ap}). A very large body of work has been
devoted to the study of population dynamics equations for more than
two species~\cite{may,drossel}. In such cases, the links among the species can
be modified according to the evolutionary dynamics. One important issue
is the control of the robustness of such models when the complexity
of the system is increased. Moreover, at a more refined
level of description, the Lotka-Volterra mean-field
dynamics can be replaced by individual-based
models~\cite{quince,md} taking into account the particularities
of the interacting individuals
and thus offering the possibility to include the stochastic fluctuations.
These dynamic food webs models allow therefore to treat on an equal footing
both the micro- and the macro-evolution of an ecosystem~\cite{chowd2,coppex}.

The richness of the models mentioned above has its own drawbacks.
Indeed, the number of control parameters defining the models is
usually quite large; moreover the dynamics is nonlinear. Thus, it is
often impossible to get a global picture of the properties of the
system. Accordingly, it is desirable to study some models which are
as simple as possible, in order to  clarify the relative importance
of the various ingredients, while being able to capture the generic
properties expected for food webs. Several proposals have been made
along this line in the past years, see e.g.~\cite{abramson,sole}. In
particular, Amaral and Meyer~\cite{amaral} proposed such a
``minimal" model whose numerical solution leads to a power-law
distribution of extinction-avalanche sizes, in good agreement with
available data from fossile record. It was shown later that this
model is self-organized critical~\cite{drossel2} and that the power
law can be obtained analytically. Furthermore, taxonomic effects
have been added to the model~\cite{CS}, but without significant
effects.

In this work we are revisiting the Amaral -- Meyer model (AM hereafter)
with the aim of investigating several of its properties  which are relevant
for real food webs and which have not been addressed in the previous works.
The paper is organized as follows. In Sec.~II, the model is described and
several technical details concerning the
Monte-Carlo simulations, as well as the values of
the control parameters, are given.
Section~III  contains the main results.
First, the dependence of the survival chance and of
the average extinction time on the number of niches $N$
and on the maximum number of feeding species $k$ is studied.
The problem of extinction due to stochastic effects
is also discussed. Then the question of the pyramidal
structure of the food web is approached.
Time correlations between the occupied niches at different
levels are investigated.
The time evolution of the ratio of omnivores is also computed,
both for viable and non-viable food webs. The distribution
of food-tree sizes as a function of the values of $N$
and $k$ is found to exhibit different regimes.
Finally, the problem of avalanches of species extinctions
is revisited.
In contradiction with previous results,
it is found that strong deviations from
simple power laws for the size distribution of
these avalanches can be observed for large values of $k$.
Some of our predictions are compared with real biological
data and are found to be in good agreement. Conclusions
are relegated to Sec.~IV.

\section{Model}

\begin{figure}
\begin{center}
\includegraphics[width=\columnwidth]{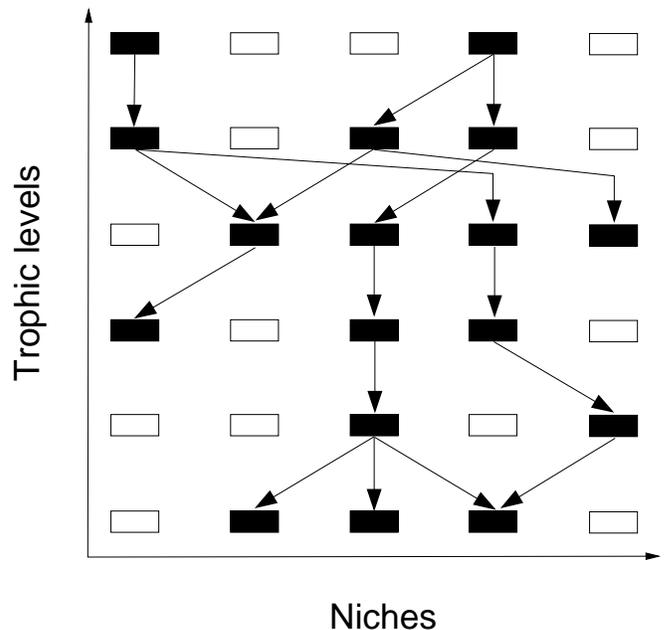}
\end{center}
\caption{Schematic representation of the AM food web model, for
$L$ = 6 and $N$ = 5. The occupied niches are represented by
the black rectangles and
the interactions between species are depicted by
the directed links.\label{fweb}}
\end{figure}

The AM food web model consists of $L$ trophic
levels, each of them containing the same number $N$
of niches, which can be either empty or occupied
by a single species. Each species from level $l=2,3, ...,L$
feeds on at most $k$ ($k\geqslant 1$) species that are randomly
selected from the level below, $(l-1)$ (see Fig.~\ref{fweb}).
Therefore a species from level $l$ is a predator
for some species at the level  $(l-1)$, and at the same time
it may be a prey for species from the level  $(l+1)$
(except for the species on the top level $L$, that
have no predators, and the species on the bottom level
$l=1$ that have no preys).

The dynamics of the web is driven by the extinction and creation of
species, as well as by the dynamically-related evolution of the
trophic links between the species. Namely, at each time step (Monte
Carlo step, MCS), one starts by randomly removing species from the
basal level 1,  with a given ``extinction probability" $p$. When a
species gets extinct, all the links from it to species at the level
2 are removed. If as the result of these link removals a species at
the level 2 looses all its preys from level 1, then it becomes
extinct as well. This procedure of checking existing links and
removal of species which lost all their food sources is then
followed on each level till the top level $L$. Hence an avalanche of
extinctions of species can be generated.

Apart from extinctions, the AM model considers also the creation of
species in the free  niches. Each species (that remains after the
decimation procedure described above) at level $l$ can repopulate,
with a probability $\mu$, an empty niche either at the level $l$,
$(l-1)$, or $(l+1)$. New-created species receive at most $k$ links,
at random, to species from the adjacent lower level.

Extinction and creation of species are thus stochastic processes
that differ from one realization of the food web to another, and one
can address the question of the statistical properties of various
characteristics of the system, like, for example, the size of the
extinction avalanches, the extinction time (or, equivalently, the
survival chance), the populations at all levels, the correlations
between the different trophic levels, the density of the trophic
links, etc. The dependence of these elements on the parameters of
the model $L$, $N$, $k$, $p$, and $\mu$ is also an important aspect
to be considered.

In this respect, the main result of the original
paper~\cite{amaral} addresing the AM model
was that the distribution of the sizes of the extinction avalanches
can be fitted over about three decades by a power law
with an exponent $a\approx -1.98$;
this exponent was corrected to the value $a=-2$ in
later works~\cite{drossel2,CS}, which is supported by mean-field theoretical
arguments. Moreover, it was argued that
this power-law behavior is in agreement with available
fossil data records. In Ref.~\cite{CS} it has also been
shown that the avalanche-size
distribution exhibits a maximum for small-size events,
before developping the power-law behavior.
However, most of the characteristics of the food web that were
enumerated above were not addressed in the previous papers
on the AM model and our work is therefore intended to fill this gap.

We shall therefore investigate the AM model in more detail, by
considering the canonical set of parameters used in~\cite{amaral},
namely $L=7$ trophic layers, the extinction probability $p$ = 0.01
and the probability of creation $\mu$ = 0.02. We shall moreover
investigate how the system characteristics depend on the number of
niches $N$ and on the highest possible number of links $k$ that a
predator may have. The obtained results will be compared to
experimental data coming from investigations of some contemporary
food webs~\cite{townsend,dunne,williams}. Since the total size --
i.e., the product of the number of layers and the mean number of
occupied niches -- of the experimentally-observed food webs does not
exceed 1000, we have decided to focus on $N$ values that are smaller
than the value of 1000 that was used in Refs.~\cite{amaral,CS}, and
to work with $N\leqslant 100$, exceptionally 200 and 500. As a
consequence, the {\em role of fluctuations} in our systems becomes
more important and many of the reported effects are clearly
noise-induced and/or noise-affected, which actually makes them more
relevant for real biological food webs. This choice of small $N$
also allowed us to run the simulations for longer times than those
considered in~\cite{amaral,CS}, which unveiled new aspects of the
food web viability. In general, we performed simulations over $\sim
10^6$ MCS and the averaging was done over 100 runs (i.e., random
realizations of the food-web stochastic dynamics). In some cases, in
order to check the viability of the system, we even went to $\sim
10^7$ MCS. Mean extinction times for the whole web were obtained by
averaging over 500 runs.


\section{Main results}

\begin{figure}
\begin{center}
\includegraphics[width=\columnwidth]{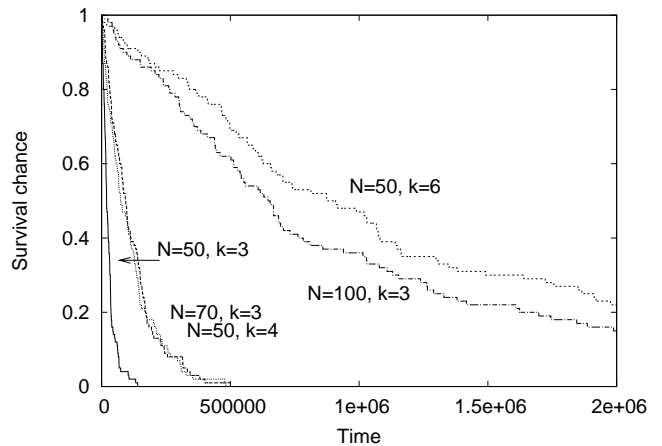}
\end{center}
\caption{Survival chance at time $t$ for different values of $N$ and $k$.
Evaluated from 100 runs.}
\label{surv}
\end{figure}

\begin{figure}
\begin{center}
\includegraphics[width=\columnwidth]{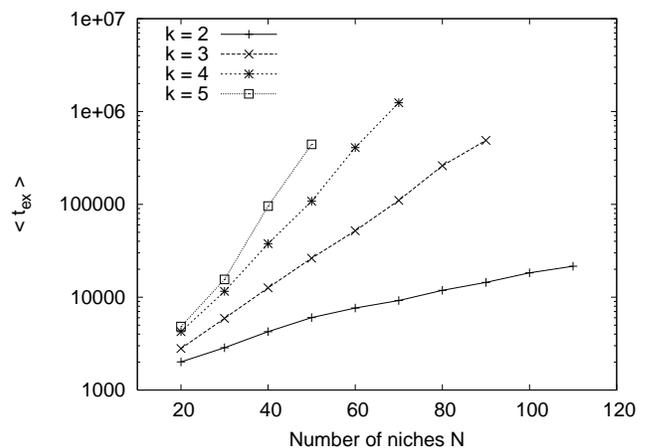}
\end{center}
\caption{Dependence of the average extinction time
of the whole food web on $N$ and $k$. The average was performed over 500 runs.}
\label{ext}
\end{figure}

A first result refers to the {\em viability} of the food web, i.e.,
to its capacity to survive in the long time limit. Performing much
longer simulations than in~\cite{amaral,CS} we have found that
small-size (e.g., $N$ = 50 or $N$ = 70) and weakly-coupled ($k$ = 3
-- 6) systems are not viable and disappear in the long time limit
$t\approx 10^6$ MCS. Figure~\ref{surv} illustrates how the chance
that a web will survive till a given time $t$ is depending on $N$
and $k$. Survival chance at time $t$ is defined here as the ratio of
the number of realizations (runs) for which the system was still
existing at time $t$, to the total number of trials.
The web of $N$ = 100 and $k$ = 3 turns out to be non-viable, too
(out of 100 runs, none has survived till $10^7$ MCS),
however increasing $k$ to 4 stabilizes the system.
The dependence of the mean extinction time on the
number of niches $N$, and on the maximum number of links $k$
is illustrated in Fig.~\ref{ext}. It is obvious that increasing $k$,
i.e., the connectedness, is stabilizing the web.
Small, sparsely coupled webs cannot exist for a longer time.

\begin{figure}
\begin{center}
\includegraphics[width=\columnwidth]{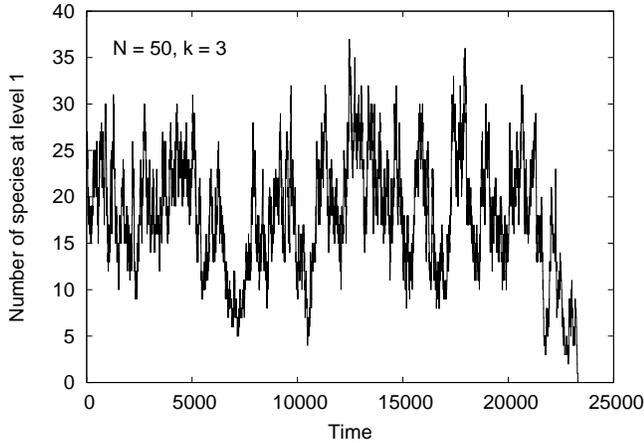}
\end{center}
\caption{Temporal evolution of the number of species on the
basal level for $N$ = 50, $k$ = 3. Single run, exhibiting the
complete extinction of the species on this level.}
\label{single}
\end{figure}

The mechanisms leading to
the collapse is connected with the stochastic nature
of the extinction and proliferation events.
Indeed, when the system is small, it may happen rather easily
that at the lowest level, which is
crucial for the survival of the web, only very few
species survive. If, moreover, as is the case in the AM model,
the values of the two creation and extinction probabilities, are very low,
then two scenarios are almost equally probable, namely:
(i) either some empty niches at the level 1 are repopulated
and the web is, at least temporarily, safe, or (ii) existing species
are all removed from this level, as illustrated in Fig.~\ref{single}.
This is the end of the food web, since without
species at the basal level an avalanche containing
all species is created and the web collapses.
This {\em stochastic extinction} in small populations
is a well known effect in ecology~\cite{shaffer}.

\begin{figure}
\begin{center}
\includegraphics[width=\columnwidth]{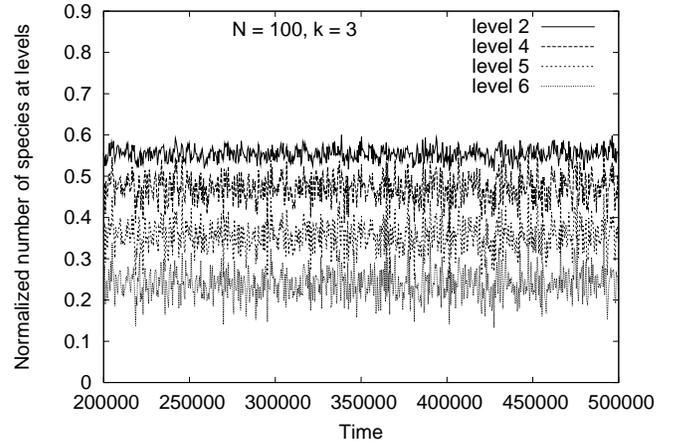}
\includegraphics[width=\columnwidth]{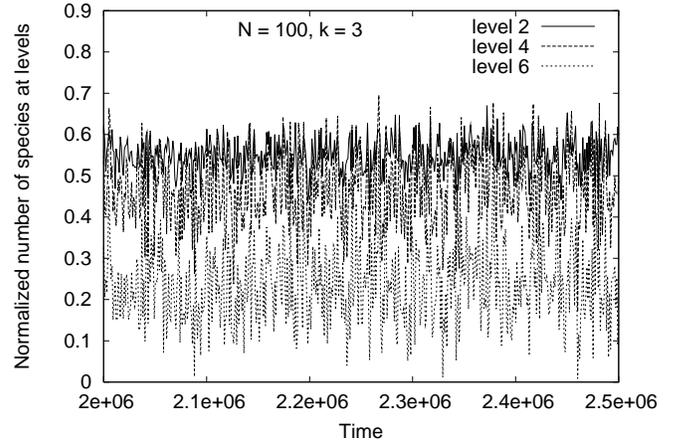}
\end{center}
\caption{Time dependence of the normalized populations of different
levels for  $N$ = 100, $k$ = 3. Upper panel: initial and mean time
stage. Lower panel: long-time stage, just before the web collapse.
Single runs.} \label{pir1}
\end{figure}

\begin{figure}
\begin{center}
\includegraphics[width=\columnwidth]{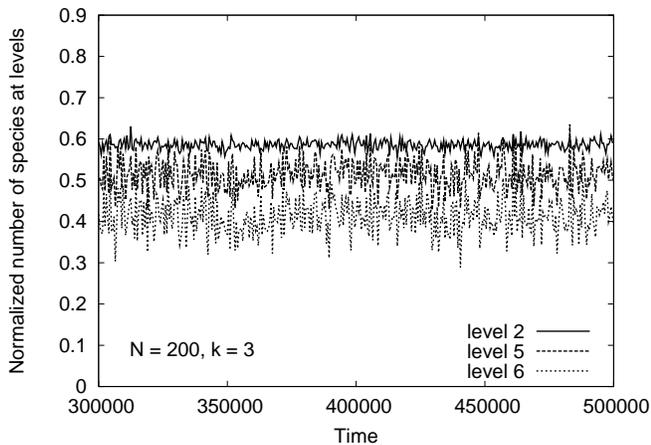}
\includegraphics[width=\columnwidth]{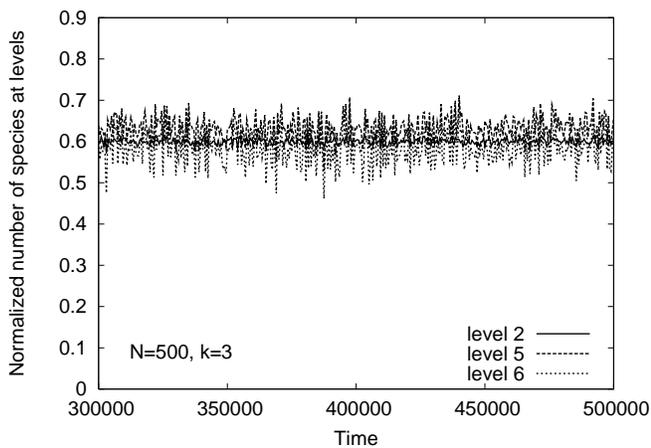}
\end{center}
\caption{Time dependence of the normalized populations of different
levels for $k$ = 3. Upper panel: $N$ = 200. Lower panel: $N$ = 500,
single runs.} \label{pir2}
\end{figure}

The next set of figures, see Figs.~\ref{pir1},  illustrate the temporal evolution of the
normalized populations at the different levels (i.e., the number of species at a given
level divided by the total number of species in the web, at a given time).
For the sake of clarity, only a part of the levels are shown.
The AM model leads in a natural way
to a {\em pyramidal form of the food web}, where the upper levels are less
populated that the lower ones, see the upper panel.
This effect is less pronounced when the system is close to its collapse,
as shown by the lower panel of Fig.~\ref{pir1}.
Comparison with Fig.~\ref{pir2}  indicates that the
pyramid-effect is practically disappearing for systems
with many niches (high $N$),
and this is the reason for which it had not been observed neither
in~\cite{amaral}, nor in~\cite{CS} for which $N$ = 1000.
Note also that the pyramidal structure has been best observed by
biologists in rather small food webs~\cite{townsend}.

As seen from both Figs.~\ref{pir1} and~\ref{pir2},
the numbers of occupied niches at different levels
are randomly oscillating in time. In order to check
the {\em degree of correlation} of these oscillations,
and whether there is some systematic time lag between them,
we have calculated the correlation functions
from the corresponding discrete-time series
of the populations on the different levels,
using the formula~\cite{orfanidis}:
\begin{equation}
C_{ij}(m) = {\cal C}_i\,\left\{
\begin{array}{ll}
\displaystyle\sum_{n=0}^{T-m}\delta N_{i}(n+m)\,\delta N_{j}(n)\,,
&\quad\mbox{for}\quad m\geqslant 0,\vspace{0.5cm}\\
\displaystyle\sum_{n=0}^{T-|m|}\delta N_{i}(n)\,\delta N_{j}(n+|m|)\,,
&\quad\mbox{for}\quad m\leqslant 0
\end{array}
\right.\nonumber
\end{equation}
Here $N_i(t)$ is the population of level $l=i$ at time $t$
(which is, of course, an integer number of MCS) and
$\delta N_i(t)$ is its fluctuation around the mean value,
\begin{equation}
\delta N_i(t)=N_i(t)-T^{-1}\displaystyle\sum_{n=0}^{T}\delta
N_{i}(n);\nonumber
\end{equation}
 $m$ is the time lag (that can be positive or negative),
and $T$ is the total simulation time. The coefficient ${\cal C}_i$
was chosen such that the autocorrelation functions at zero lag are
equal to 1,
\begin{equation}
{\cal C}_i= \left[\displaystyle\sum_{n=0}^{T}(\delta
N_{i}(n))^2\right]^{-1}.\nonumber
\end{equation}
 The results for $N=100$ and $N=200$ are
illustrated in Figs.~\ref{correlations, n100} and \ref{correlations,
n200}, respectively.
\begin{figure}
\begin{center}
\includegraphics[width=\columnwidth]{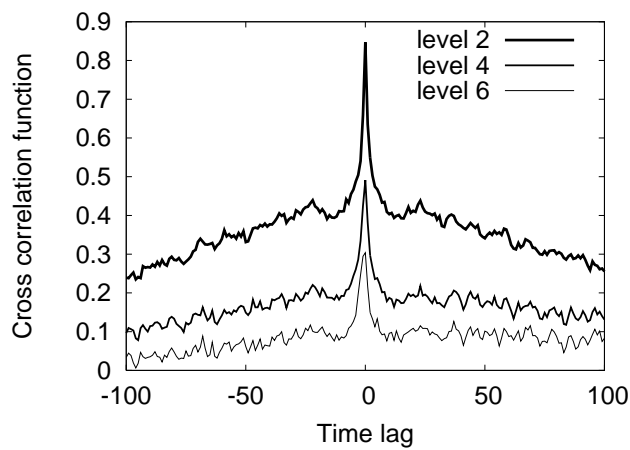}
\caption{Cross-correlations between the occupation numbers at levels
$i=1$ and $j=2,4,6$. Here $N=100$ and $k=3$, and the evaluations are
done on a single time-series. \label{correlations, n100}}
\end{center}
\end{figure}
\begin{figure}
\begin{center}
\includegraphics[width=0.45\textwidth]{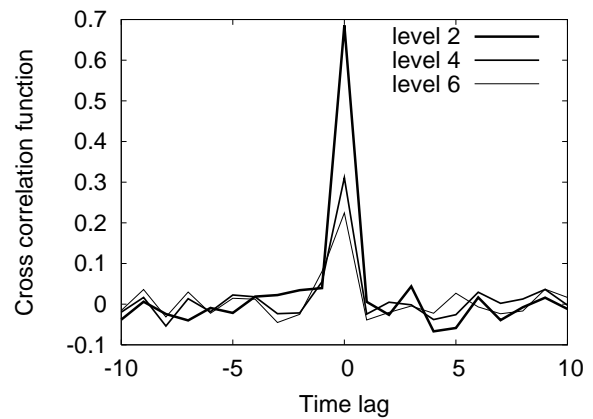}
\caption{Cross-correlations between the occupation numbers at levels
$i=1$ and $j=2,4,6$. $N=200$, $k=3$, evaluations from single runs.
\label{correlations, n200}}
\end{center}
\end{figure}
From these figures it follows that the
time series at neighboring levels are highly
correlated with each other at zero time lag.
In other words, species at a given level adjust
immediately to the changes at the level below,
which is a feature that could be expected in view
of the constitutive dynamics of the model.
The correlation is of course decreasing with the distance
between the levels, but the peak at zero time lag remains.
It should be noted that in the $N=100$ case, Fig.~\ref{correlations, n100},
apart from the very narrow zero-lag correlation peak
there is also a rather broad structure centered around it.
This structure is practically absent in the case of the larger system
with $N=200$, see Fig.~\ref{correlations, n200} and is to
be related to the long-time instability of the system with  $N=100$ and $k=3$,
and to the strong fluctuations that are accompanying its collapse.

\begin{figure}
\begin{center}
\includegraphics[width=\columnwidth]{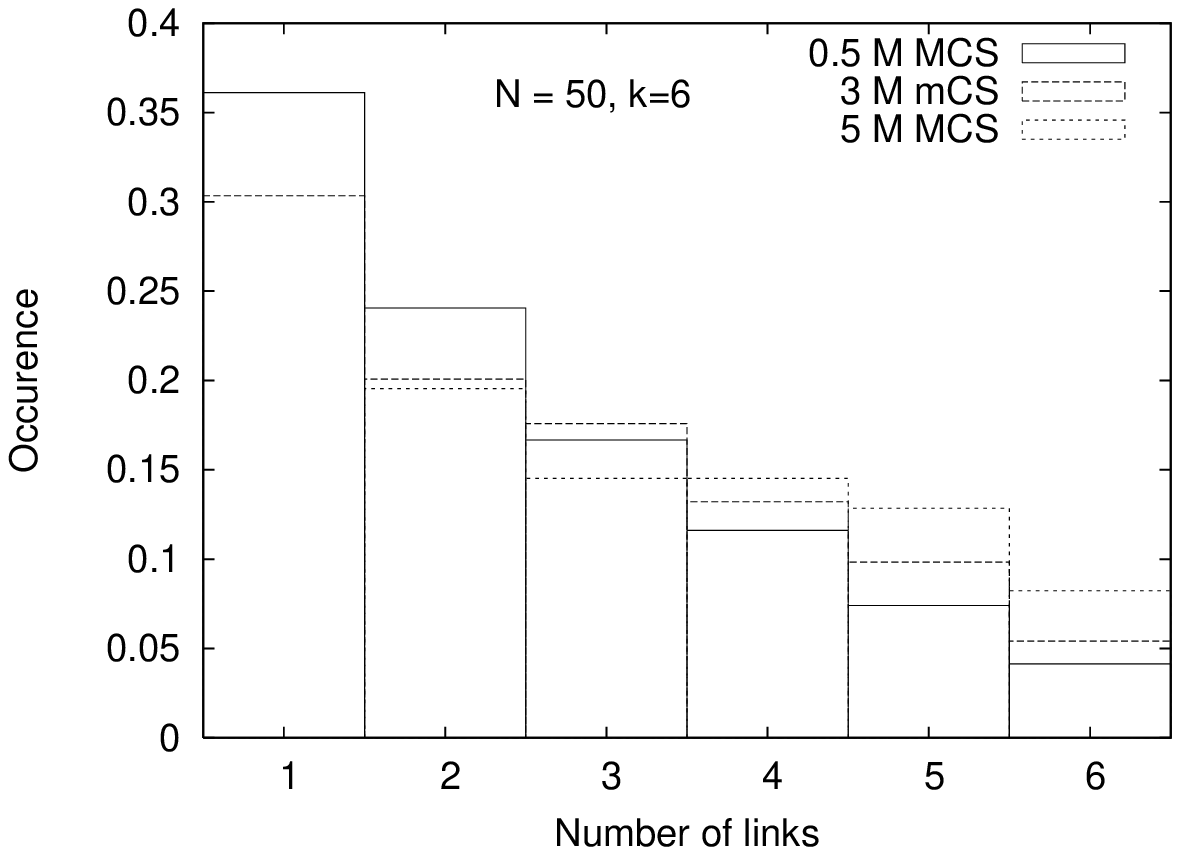}
\includegraphics[width=\columnwidth]{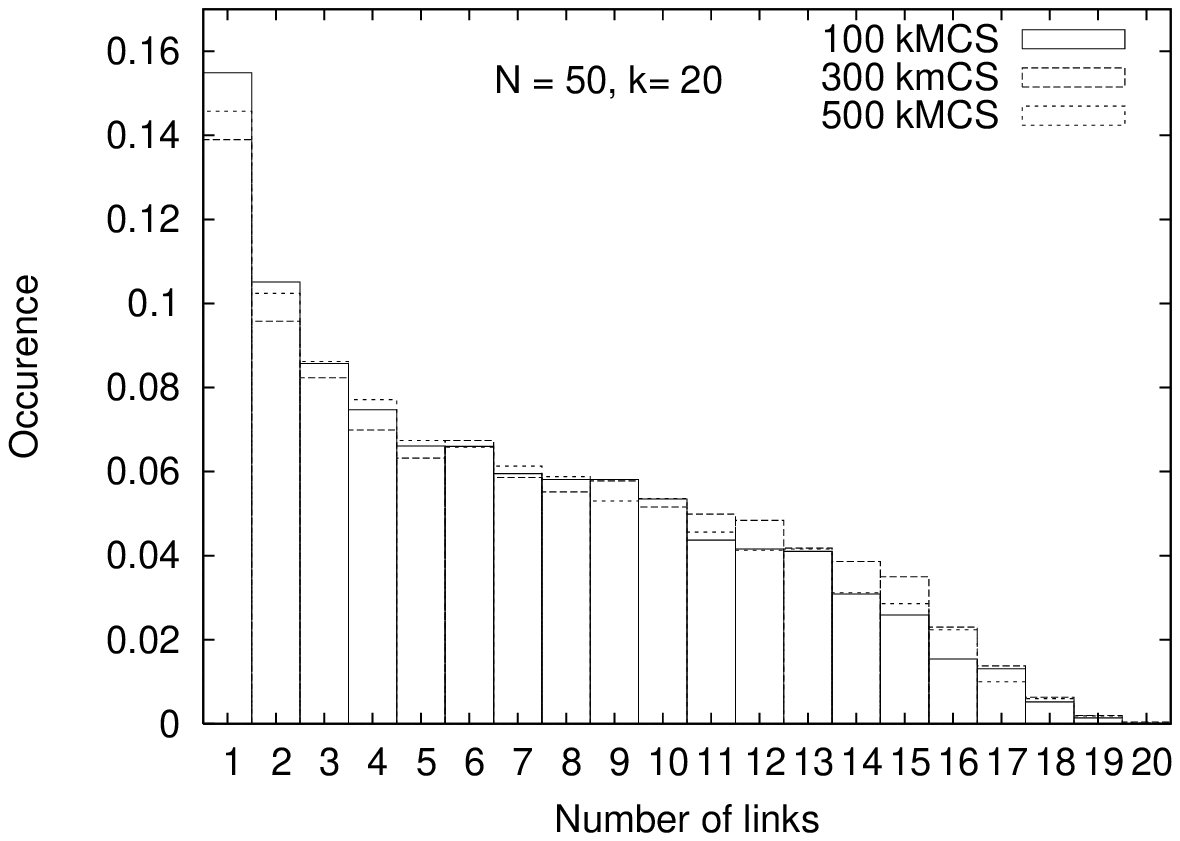}
\end{center}
\caption{ Distribution of the number of links per predator for
non-viable  $k$ = 6 (upper plot), and viable $k$ = 20 (lower plot)
systems, at different times. $N$ = 50 in both plots, evaluations
from single runs.} \label{dlink}
\end{figure}

Another biologically interesting feature is the {\em fraction of
omnivores}, that are predators feeding on more than one
prey~\cite{borrval}.  Figures~\ref{dlink} shows the distribution of
the number of links per predator for a non-viable (upper panel), and
a viable (lower panel) system, at several times. The distribution
remains virtually the same throughout the simulation time for a
viable web. However, for a non-viable one approaching extinction,
the fraction of highly-connected predators grows. In other words,
close to the collapse, only predators feeding on many preys will
survive. In this sense, the presence of omnivores stabilizes the
web, as documented experimentally in~\cite{borrval}.

The average fraction of omnivores in a stationary
state of the food web depends on the maximum value
of links $k$, but it is rather insensitive to the number
of niches $N$, as illustrated by the data in
Table~\ref{tab1}. This feature also agrees with experimental
results reported in~\cite{williams}.

\begin{table}
\begin{center}
\begin{tabular}{l|c|c|c|c}\hline\hline
\backslashbox[0pt][l]{$N$}{$k$} & 3&4&6&20\\ \hline
50&0.312&0.4185&0.6966&0.8543 \\
100&0.2936&0.4944&0.6567&0.8648\\
\hline\hline
\end{tabular}
\caption{Average fraction of omnivores in a stationary state of the web, for various $k$
and $N$. The average was taken over 100 runs.
\label{tab1}}
\end{center}
\end{table}

\begin{figure}
\begin{center}
\includegraphics[width=\columnwidth]{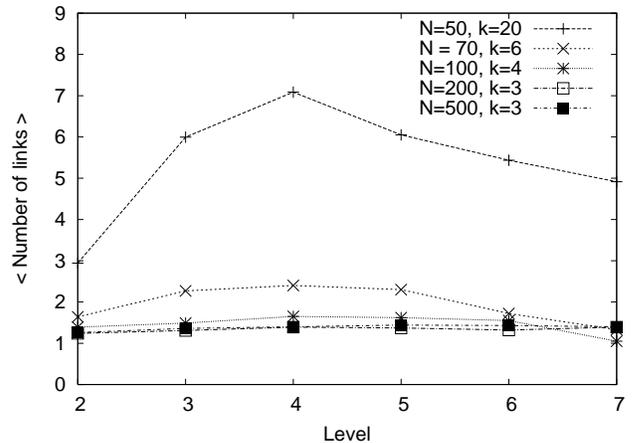}
\end{center}
\caption{ Number of links for different levels in a stationary state
of the web, averaged over 100 runs.} \label{nlinks}
\end{figure}

In a stationary state of the web,
the average number of links corresponding to different levels $l$ has
the same type of profile whatever the value of $k$, namely a more or less
pronounced maximum for the intermediary levels and a drop for the
low and top levels, as illustrated in Fig.~\ref{nlinks}.

Food webs are also often
characterized by the {\em length of the food chains} (or ``trees")
that are forming the web~\cite{townsend,williams}. We define them
in the following way: each species with no predators
is the root of a new tree. Starting from the root we go
along its links to the lower level and mark all species
the root is predating on. Then we check their links to find their
prey species and so on.  Since different predators do not
really compete for food in the model (i.e., if they are linked
to the same prey, they all get enough food),
we can treat the partially overlapping trees as independent ones.
The size of a tree is then simply the total number of species that
belong to that tree.

As can be seen from Fig.~\ref{tree, t300k}, food tree size
distribution depends on the number of niches $N$ in the system.
The maximal tree size increases with $N$, as could be expected.
Moreover, in a bigger system there is more space for trees of
similar sizes and that is why the curves in Fig.~\ref{tree, t300k}
shift upwards with increasing  $N$.  Linear dependence of
the distribution of chain lengths on the semi-logarithmic
plot in Fig.~\ref{tree, t300k} indicates
an exponential decrease with the tree size.
\begin{figure}
\begin{center}
\includegraphics[scale=0.7]{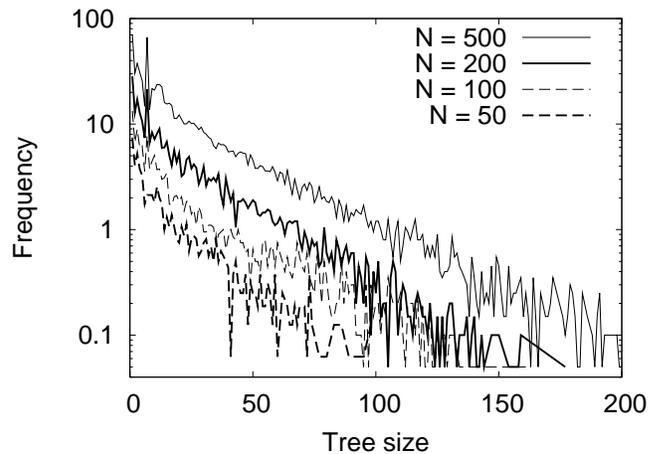}
\caption{ Food tree size distribution at time $T=3\, \cdot\,
10^{5}~MCS$ as a function of the number of niches $N$.  $k=6$,
semi-logarithmic scale, evaluation from 100 runs. \label{tree,
t300k}}
\end{center}
\end{figure}
\begin{figure}
\begin{center}
\includegraphics[width=0.45\textwidth]{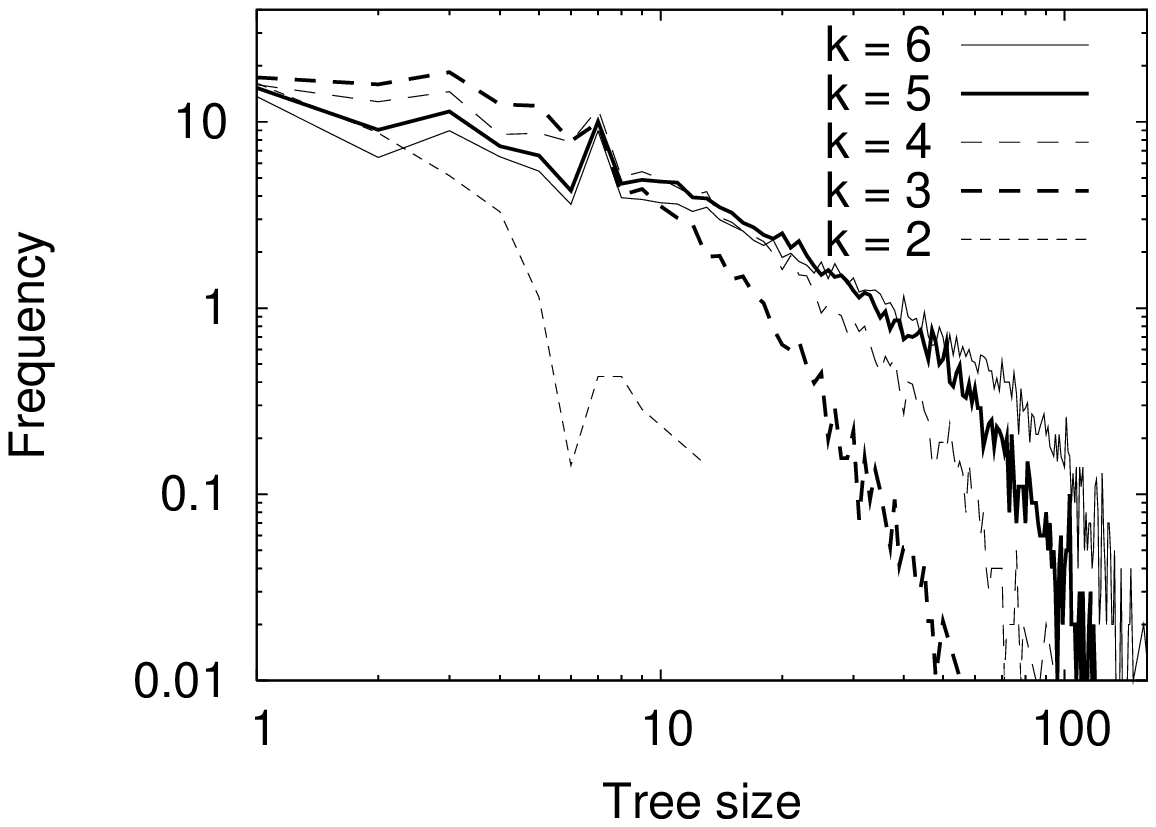}
\includegraphics[width=0.45\textwidth]{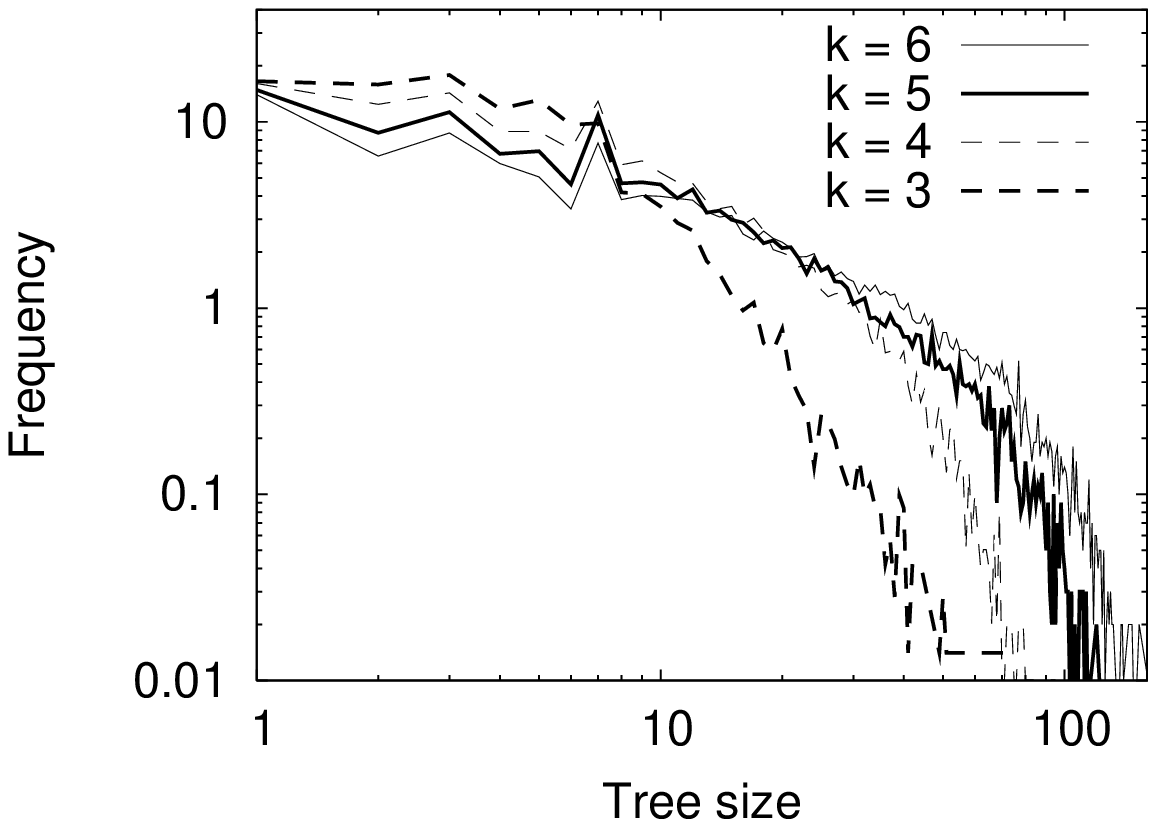}
\caption{ Food tree size distribution at time
$T=5\,\cdot\,10^{4}~MCS$ (upper plot) and $T=3\,\cdot\,10^{5}~MCS$
(lower plot) as a function of the maximal number of links between
species $k$.  $N=100$, log-log scale, evaluation from 100 runs.
\label{tree, k}}
\end{center}
\end{figure}

The maximum number of links $k$ between species is
also playing an important role on the food tree size distribution.
The results for $N=100$ niches at two  different time steps
are displayed in Fig.~\ref{tree, k}.
When $k$ increases, small trees become less likely and bigger
structures in the system are preferred instead. We can thus
distinguish two regimes with different $k$-dependence,
namely the regime of ``small trees" (of  size $\leqslant 10$),
whose number decreases with $k$, while the number of
``big trees" is an increasing function of $k$.
Moreover, the maximal size of a food tree
varies strongly with $k$.  It is also interesting to note
that for sufficiently large $N$ there is a well-pronounced peak in the
distribution of tree sizes at a size of $7$, which  is
simply the number of trophic levels in the system.

Biologists~\cite{paine,williams} often describe food webs
in terms of fractions of {\it basal, intermediate} and
{\it top} species. In this model these ones correspond  to occupied
niches at levels 1, 2--6, and 7, respectively. For viable
systems the values we obtain for these fractions are not too sensitive to
the values of $N$
and are presented in Table~\ref{tab2}.
\begin{table}
\begin{center}
\begin{tabular}{l|c|c|c}
\hline\hline
N& basal&interm.&top\\ \hline
100&0.16&0.79&0.05\\
200& 0.10 & 0.80 & 0.10 \\
500 & 0.12 & 0.79 & 0.09\\
\hline\hline
\end{tabular}
\caption{Fraction of species at basal, intermediary and top
levels. $k=6$, average over 100 runs.\label{tab2}}
\end{center}
\end{table}
These results  agree very well with  biological data for
a food web from Little Rock, U.S.A., see~\cite{williams} for further details.

\begin{figure}
\begin{center}
\includegraphics[width=\columnwidth]{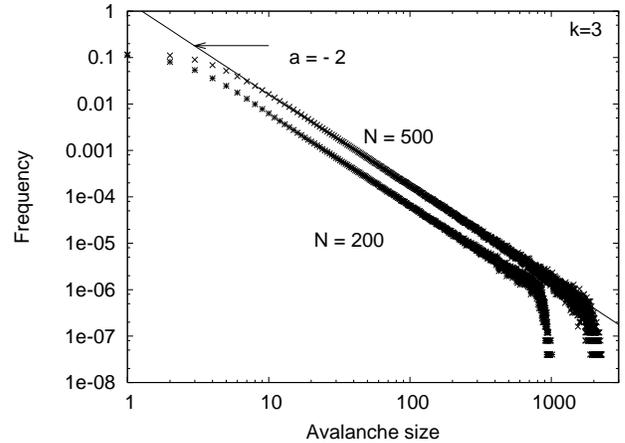}
\end{center}
\caption{Distribution of avalanche sizes for $N$ = 200 and $N$ = 500, for food webs
with $k=3$. Estimated from 500 runs.}
\label{aval500}
\end{figure}

\begin{figure}
\begin{center}
\includegraphics[width=\columnwidth]{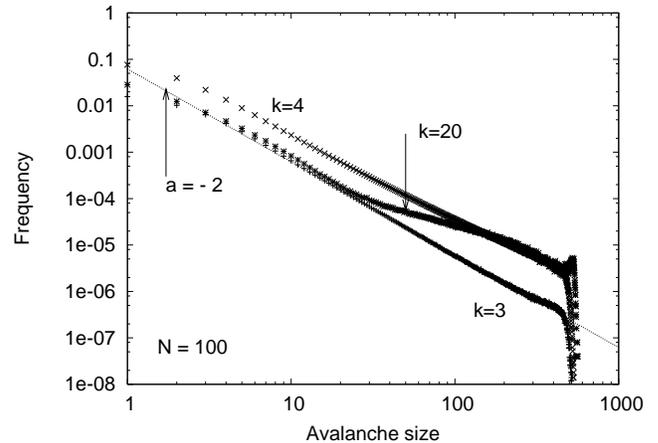}
\end{center}
\caption{Distribution of avalanches for $N$ = 100 and different $k$.
Estimated from 500 runs.}
\label{aval100}
\end{figure}

Finally, we have analyzed the distribution of {\em avalanche sizes
of species extinctions}. We have observed the maximum in the
distribution that was mentioned in~\cite{CS}, which becomes more
pronounced with increasing the number $N$ of niches. For $k$ = 3,
which was the value considered in~\cite{amaral,CS}, we recovered the
known power-law behavior, extending over nearly three decades, with
an exponent equal to $a=- 2$, as calculated in~\cite{drossel2,CS}.
This value does not seem to depend on $N$, and even for non-viable
systems we got the same good fit to a power-law type of behavior,
with the same exponent, see Figs.~\ref{aval500}, ~\ref{aval100}.

\begin{figure}
\begin{center}
\includegraphics[width=\columnwidth]{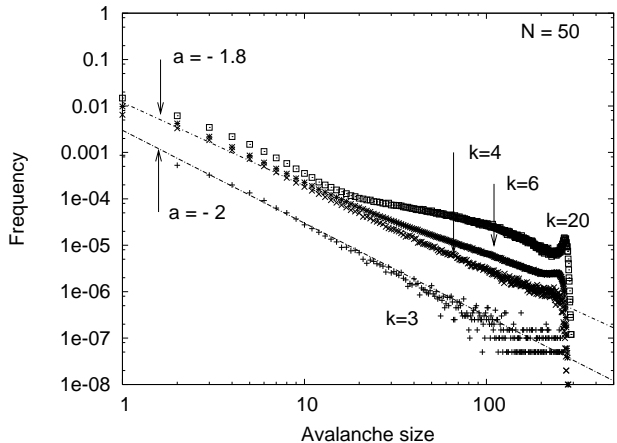}
\end{center}
\caption{Distribution of avalanches for $N$ = 50
and different $k$. Estimated from 500 runs.}
\label{aval50}
\end{figure}

However, when the food web becomes highly connected (i.e., $k$ = 6
or larger) the deviations from the power-law behavior are very
large, as illustrated in Figs.~\ref{aval100},~\ref{aval50} for $k$ =
20. One may notice that for highly-connected webs the fraction of
larger avalanches increases, simply because the removal of a prey on
which many predators feed is affecting more species.

As such, the power-law behavior, although largely
present in the large-$N$ webs, is {\em not universal},
but it is rather the result
of a particular choice of the parameter values of the food web.


\section{Conclusions}

We have presented a detailed discussion of several aspects of
the AM model~\cite{amaral} of a food web that were neglected
in the previous studies. In particular, we concentrated on relatively
small-size systems and on the role the fluctuations can play in
such systems, since this is the frame  that is important
in most of the real biological food webs. Several comparisons of the
theoretical predictions  with experimental data were also discussed.

Our simulations confirmed the observations of~\cite{CS} concerning
the distribution of avalanche sizes of species extinctions,
the value of the power-law exponent $a=-2$, and the existence
of a maximum depending on $N$. It is worth noting here that the same value
$a=-2$ of the exponent of the distribution of extinction avalanches
has also been found  in two other different food web models,
Refs.~\cite{rikvold} and ~\cite{roberts}.  The last paper
is a generalization of the Bak and Sneppen
model ~\cite{bs}, in which two factors determine the fate
of a species -- biotic (``bad genes") and abiotic (``bad luck").

For reference, we have kept the values of some of the model
parameters (number of layers $L$, probability of creation $\mu$ and
of extinction $p$) the same as in the original AM
model~\cite{amaral,CS}. However the dependence of the system
behavior on the number of niches, $N$ and on the maximum number of
links per predator $k$, turned out to be quite interesting. We have
thus unveiled new features of the model, not found in the earlier
papers. A food web may collapse if it is too small and/or has not
enough links between species. Systems smaller than $N \approx  200$
show a pyramid-like structure, where top levels are less populated
than the bottom ones. The occupations of the levels are strongly
correlated at zero time lag. When the web is close to a collapse,
the fraction of highly connected predators (omnivores) significantly
increases, which may lead sometimes to a (temporary) rescue of the
web. The distribution of the length of  food trees has an
exponential character and its type depends rather weakly on both $N$
and $k$. Finally, the distribution of species extinctions shows an
unexpected feature, contrary to the previously-claimed universality
of the power-law behavior; namely that for large $k$-values (i.e.,
highly-connected webs) the distribution cannot be described anymore
as a power law.

Although several criticisms concerning the applicability of the AM
model to biology have been raised~\cite{newman,CS}, we have found
that some theoretical results, like the ratio of omnivores, the
fraction of different-type (basal, middle, top) species, the food
chain length etc.,  are in very good agreement  with experimental
data~\cite{townsend,dunne,williams} on food webs. The significance
of this fact is a subject of further analysis.

\acknowledgments
This work has been done within the UNESCO Chair for
Interdisciplinary Studies activity.
AP and JS acknowledge financial support from COST P10 Project
``Physics of Risk''  and the hospitality
of the D\'epartement de Physique Th\'eorique
de l'Universit\'e de Gen\`eve. This work was partly supported by the Swiss
National Science Foundation.

\end{document}